\documentclass[epj]{svjour}
\usepackage{psfig}
\begin{document}
\title{First observation of $^{55,56}$Zn}

\author{J.~Giovinazzo\inst{1} \and B.~Blank\inst{1} \and C.~Borcea\inst{2} \and
M.~Chartier\inst{1} \and S.~Czajkowski\inst{1} \and
G.~de~France\inst{3} \and 
R.~Grzywacz\inst{4}\thanks{\emph{Present address:
Univ. of Tennessee, Knoxville, Tennessee, USA}} \and Z.~Janas\inst{4} \and  M.~Lewitowicz\inst{3} \and 
F.~de~Oliveira Santos\inst{3} \and M.~Pf\"utzner\inst{4} \and M.S.~Pravikoff\inst{1} \and  J.C.~Thomas\inst{1}}

\institute{Centre d'Etudes Nucl\'eaires de Bordeaux-Gradignan, Le Haut-Vigneau, 
B.P. 120, F-33175 Gradignan Cedex, France
\and
Institute of Atomic Physics, P.O. Box MG6, Bucharest-Margurele, Romania
\and 
Grand Acc\'el\'erateur National d'Ions Lourds, B.P. 5027, F-14076 Caen Cedex, France
\and
Institute of Experimental Physics, University of Warsaw, PL-00-681 Warsaw, Hoza 69, Poland
}

\date{Received: date / Revised version: date}

\abstract{
In an experiment at the SISSI/LISE3 facility of GANIL,
the most proton-rich zinc isotopes $^{55,56}$Zn have been observed 
for the first time. The experiment was performed using a high-intensity
$^{58}$Ni beam at 74.5~MeV/nucleon impinging on a nickel target.
The identification of $^{55,56}$Zn opens the way to $^{54}$Zn, 
a good candidate for two-proton radioactivity according to theoretical predictions.
\PACS{27.40.+z \and 21.10.Dr \and 23.50.+z \and 25.70.Mn
} % end of PACS codes
} %end of abstract

\maketitle

Nuclear structure experiments near the proton drip line represent an
important tool to investigate the properties of the atomic nucleus. The
mapping of the proton drip line provides a first stringent test for mass
models. One of the most exciting new phenomena at the proton drip line 
is probably the occurence of the two-proton ground-state (2p) decay which has 
been predicted about 40 years ago~\cite{goldanskii60}. Although
considerable experimental and theoretical efforts have been made in order 
to observe this new radioactivity, no evidence was found up to now. 
Instead, a three-body break-up regime has been observed in the ground-state 
decays of $^6$Be~\cite{bochkarev92} and $^{12}$O~\cite{kryger95}. In 
$\beta$-delayed processes where the emission occurs from
excited states, a sequential-emission picture via an intermediate state 
is able to describe the experimental data (see e.g.~\cite{cable84,axelsson98}.

For the 2p-decay candidates, the two-proton separation energy is 
negative. In the case of nuclei, for which one-proton emission is energetically 
forbidden, the emission of the two protons has to be simultanous. Therefore,
there are two possibilities: i) an uncorrelated emission where the two protons
occupy the whole phase space or ii) a correlated emission where an angular
correlation of the two protons may be observed. According to mass
predictions~\cite{brown91,ormand96,cole96,ormand97}, the best candidates for 
this radioactivity are $^{45}$Fe, $^{48}$Ni, and $^{54}$Zn
with predicted half lives in the 1$\mu$s - 1ms range.

$^{45}$Fe and $^{48}$Ni have been observed in recent experiments at 
GSI~\cite{blank96fe45} and at GANIL~\cite{blank00ni48}. The production rate
of $^{45}$Fe at the LISE3 separator~\cite{lise} of GANIL reaches now a level
where spectroscopic studies become feasible to investigate the principal
decay modes of the exotic proton-rich nuclei. However, $^{48}$Ni is produced
only with 1-2 counts per day, making spectroscopic studies difficult, if not
impossible. Therefore, it seems to be very interesting to search for new
candidates for this decay mode. In the lighter mass region, the proton
drip line has been reached for all even-Z nuclei up to nickel. The search 
for new 2p candidates has thus to concentrate on the medium-mass region above 
nickel.

The next heavier even-Z element being zinc, we sear\-ched for $^{55,56}$Zn 
during the run leading to the discovery of doubly-magic $^{48}$Ni. These
isotopes were in the acceptance of the LISE3 separator for an intermediate 
setting on $^{52}$Ni used to optimize the SISSI and LISE3 settings for
the search of $^{48}$Ni. The estimated transmissions were 8\% and 0.5\%
for $^{55}$Zn and $^{56}$Zn, respectively. However, as the transmission
of LISE3 was not optimized for these nuclei, $^{56}$Zn ions were 
transmitted at the edge of the acceptance of LISE3, and these calculations
may have therefore an error of about a factor of two.

The fragments of interest were produced by projectile fragmentation of a 
primary $^{58}$Ni$^{26+}$ beam with an intensity of about 1$\mu$A and an energy 
of 74.5~MeV/nucleon. This beam was fragmented on a 230.6~mg/cm$^2$ thick
natural nickel target followed by a 2.7~mg/cm$^2$ carbon stripper foil.
After passing the first LISE dipole section, the selected fragments impinged
on a 10.36~mg/cm$^2$ beryllium degrader before entering the second dipole
stage and the velocity filter. 

In this part of the experiment, we used a three-element implantation device 
consisting of three silicon detectors with thicknesses of 300$\mu$m, 700$\mu$m, 
and 6~mm. The nuclides of interest were stopped in the second detector.
Two time-of-flight (TOF) signals were measured between 
i) a micro-channelplate (MCP)
detector at the exit of the second LISE dipole stage and the implantation
setup (TOF1) as well as ii) between the cyclotron radiofrequency and the 
implantation setup (TOF2). All parameters were calibrated by means of the 
primary beam.
Fragments transmitted to the end of the LISE3 separator were identified
event-by-event using five parameters: i) the energy loss in the first
silicon detector, ii) the residual energy in the second detector, iii) TOF1,
iv) TOF2, and v) the veto signal from the third silicon detector.

\begin{figure}[h]
\psfig{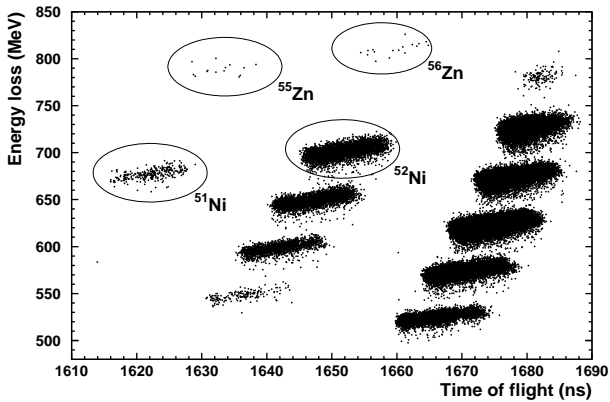}
\caption{Two-dimensional identification plot of energy loss in the first 
         silicon detector versus the TOF between the target and the 
         silicon detector generated by means of the cyclotron radiofrequency. 
         To produce this spectrum, software conditions were applied to all
         parameters except those represented here. 
         }
\protect\label{zn55_56}
\end{figure}

As in the analysis of the runs optimized on $^{48}$Ni, the position and the
width of each nuclide have been determined for each of the five parameters.
In order to accept a count as a valid event, each value of the parameters
(except those used for the representation, see Fig.~\ref{zn55_56})
had to lie within a two-FWHM window of the determined central position.
The results of this analysis is presented in Fig.~\ref{zn55_56}. We observe 
17~counts of $^{56}$Zn and 14~counts of $^{55}$Zn. Although all isotopes,
even those known to be unbound like $^{53,54}$Cu, have been analysed in the same
way and thus admitted to show up in the spectrum, almost no background events 
are visible in the spectrum of Fig.~\ref{zn55_56}. The isotope identification
is unambiguous, as i) the limits of stability in the mass region are very 
well known and ii) all parameters were calibrated at the beginning of the experiment with the primary beam transmitted to the final LISE3 focus.
These counting rates, together with the estimated transmissions, lead
to production cross sections of about 2$^{+0.6}_{-0.5}\times$10$^{-8}$mb 
for $^{55}$Zn and of 5$^{+20}_{~-2}\times$10$^{-7}$mb for $^{56}$Zn.

The nucleus $^{56}$Zn is predicted bound by all commonly used mass 
models~\cite{ormand97,audi95,moeller95,aboussir95,jaenecke88,pape88} 
with a two-proton separation energy ranging from 0.1~MeV to 1.2~MeV. 
This nuclide is expected to decay by $\beta$ and $\beta$p decay to 
either $^{56}$Cu or $^{55}$Ni. 

$^{55}$Zn is also expected to $\beta^+$-decay to states in $^{55}$Cu
which may then decay by one- or two-proton emission to $^{54}$Ni or $^{53}$Co,
respectively. However, some models~\cite{ormand97,moeller95,jaenecke88,pape88}
predict a negative two-proton separation energy ranging between -0.4~MeV and 
-1.2~MeV. It is interesting to note that a two-proton emission Q value
of 1.2~MeV yields a barrier-penetration half-life of about 10~ms in the
di-proton model which assumes that the two protons are in a relative $s$ 
state with zero binding energy.

From the experimental observation of $^{55,56}$Zn and the counting rates 
measured in the present experiment, extrapolations for an experiment optimized
for the identification of $^{54}$Zn and for the observation of its main decay
channels are possible. Assuming similar conditions as in the present experiment,
however, with the LISE3 separator optimized on $^{54}$Zn, we expect about 
10-20 events for $^{54}$Zn per day for a primary-beam intensity of about 
3~$\mu$A. This production rate should be sufficient to determine the principal
decay branches of this nuclide. It is hoped that $^{54}$Zn decays by
2p emission, as all commonly used mass 
predictions~\cite{ormand97,audi95,moeller95,aboussir95,jaenecke88,pape88} 
expect it to be two-proton unbound.
Except the 2p separation energy of Aboussir {\it et al.}~\cite{aboussir95}
which yields a value of only -0.1~MeV, all other models give values between
-1.5~MeV and -2.2~MeV.

In summary, we observed for the first time the proton-rich zinc isotopes
$^{55,56}$Zn in a projectile-fragmentation experiment at the LISE3 facility of
GANIL. The observed production rates indicate that the two-proton emission
candidate $^{54}$Zn may be produced with a rate of 10-20 events per day
in an experiment optimized for its production.

We would like to acknowledge the continous effort of the GANIL  
staff for ensuring a smooth 
running of the accelerators and the LISE3 separator. This work was 
supported in part by the Polish Committee of Scientific Research under
grant KBN 2 P03B 036 15, the 
contract between IN2P3 and Poland, as well as by the Conseil R\'egional 
d'Aquitaine.


\begin{thebibliography}{10}

\bibitem{goldanskii60}
V.~I. Goldanskii, Nucl. Phys. {\bf 19},  482  (1960).

\bibitem{bochkarev92}
O.~V. Bochkarev {\it et~al.}, Sov. J. Nucl. Phys. {\bf 55},  955  (1992).

\bibitem{kryger95}
R.~A. Kryger {\it et~al.}, Phys. Rev. Lett. {\bf 74},  860  (1995).

\bibitem{cable84}
M.~D. Cable {\it et~al.}, Phys. Rev. C {\bf 30},  1276  (1984).

\bibitem{axelsson98}
L. Axelsson {\it et~al.}, Nucl. Phys. A {\bf 628},  345  (1998).

\bibitem{brown91}
B.~A. Brown, Phys. Rev. C {\bf 43},  R1513  (1991).

\bibitem{ormand96}
W.~E. Ormand, Phys. Rev. C {\bf 53},  214  (1996).

\bibitem{cole96}
B.~J. Cole, Phys. Rev. C {\bf 54},  1240  (1996).

\bibitem{ormand97}
W.~E. Ormand, Phys. Rev. C {\bf 55},  2407  (1997).

\bibitem{blank96fe45}
B. Blank {\it et~al.}, Phys. Rev. Lett. {\bf 77},  2893  (1996).

\bibitem{blank00ni48}
B. Blank {\it et~al.}, Phys. Rev. Lett. {\bf 84},  1116  (2000).

\bibitem{lise}
A.~C. Mueller and R. Anne, Nucl. Instrum. Meth. {\bf B56},  559  (1991).

\bibitem{audi95}
G. Audi and A.~H. Wapstra, Nucl. Phys. {\bf A595},  409  (1995).

\bibitem{moeller95}
P. M{\"o}ller, J.~R. Nix, W.~D. Myers, and W.~J. Swiatecki, At. Data Nucl. Data
  Tab. {\bf 59},  185  (1995).

\bibitem{aboussir95}
Y. Aboussir, J. Pearson, A. Dutta, and F. Tondeur, At. Data Nucl. Data Tab.
  {\bf 61},  127  (1995).

\bibitem{jaenecke88}
J. J{\"a}necke and P. Masson, At. Data Nucl. Data Tab. {\bf 39},  265  (1988).

\bibitem{pape88}
A. Pape and M. Antony, At. Data Nucl. Data Tab. {\bf 39},  201  (1988).

\end{thebibliography}
\end{document}